\documentclass[letter, 10pt]{article}

\usepackage[margin=1in]{geometry}
\usepackage[numbers,sort&compress]{natbib} 
\usepackage{authblk} 
\usepackage{color}
\usepackage{amsmath}
\usepackage{amssymb} 
\usepackage{graphicx} 
\usepackage{fancyhdr} 
\usepackage{amsfonts}
\usepackage{tabularx} 
\usepackage{slashed}
\usepackage[
singlelinecheck=false 
]{caption}

\def\mpl{\text{M}_\text{pl}}
\def\Sb{\bar{S}}
\def\be{\begin{equation}}
\def\ee{\end{equation}}
\def\bea{\begin{eqnarray}}
\def\eea{\end{eqnarray}}
\def\half{\frac{1}{2}}
\newcommand{\lrs}[1]{\left[{#1}\right]}
\newcommand{\lrp}[1]{\left({#1}\right)}
\newcommand\zfr{$\mathbb{Z}_4^R$~}
\newcommand\zfrmm{\mathbb{Z}_4^R~}
\newcommand{\fw}{\ensuremath{\mathcal{W}}}
\newcommand{\fk}{\ensuremath{\mathcal{K}}}

\newcommand{\fo}{\ensuremath{\mathcal{O}}}

\begin{document}

\renewcommand\headrule{} 

\title{Leptogenesis after Inflation in a Pati-Salam Model}
\author{Stuart Raby}
\affil{\emph{Department of Physics}\\\emph{The Ohio State University}\\\emph{191 W.~Woodruff Ave, Columbus, OH 43210, USA}}

\maketitle
\thispagestyle{fancy}

\begin{abstract}
In this talk I discuss a supersymmetric Pati-Salam model of fermion masses and mixing angles which fits
low energy data.   The model is then extended to include an inflationary sector which is shown to be consistent
with Bicep2-Keck-Planck data.  The energy scale during inflation is associated with the PS symmetry breaking scale.
Finally, the model is shown to be consistent with the observed baryon-to-entropy ratio necessary for Big Bang Nucleosynthesis.
It turns out that only the heaviest right-handed neutrino decays produce the correct sign of the baryon-to-entropy ratio.
Nevertheless, we obtain the observed value due to the process of instant preheating.
\end{abstract}

\pagenumbering{gobble}

\section{INTRODUCTION}
A grand unified theory must be able to fit low energy data.  In addition, it must be able to consistently describe
the early universe.   In a recent paper \cite{Bryant:2016tzg} we have described a Pati-Salam model with a $D_4$ family
symmetry which fits low energy data quite well (see \cite{Anandakrishnan:2012tj,Anandakrishnan:2014nea,Poh:2015wta} which is based
on the model introduced in \cite{Dermisek:2005ij}).  The biggest discrepancy is in the fits for the up and down quark masses
which are too large.\footnote{We have found that we can add one new complex parameter to the Yukawa matrices, obtaining a smaller $\chi^2/dof$ fit to
low energy data.  We have checked that the results discussed in this talk are unaffected.} Bicep2, Keck and Planck cosmological data are consistent with inflationary early universe with an energy
density during inflation of order a GUT scale,  $10^{16}$ GeV.  In the Pati-Salam [PS] model, this is the scale where Pati-Salam is
broken to the Standard Model gauge symmetry in a process called ``hybrid inflation."   It is ``subcritical hybrid inflation" since
inflation begins after the waterfall field begins to slide down the side of the potential.   Since the waterfall field
breaks the PS symmetry, any monopoles formed will be severely diluted during inflation.

Of course,  after inflation the universe reheats and at this time a baryon asymmetry must be generated.   It is the process
of baryogenesis via leptogenesis which we consider now.   But first let us briefly review the PS model and the results of subcritical
hybrid inflation.

\section{Pati-Salam 3 family model with $D_4$ family symmetry}

The matter sector of the theory is given by the superpotential $\fw = \fw_I + \fw_{PS}$ with
\begin{align}
  \begin{aligned}
    \fw_{PS} =&\, \fw_{neutrino} + \lambda Q_3 {\cal H} Q^c_3 + Q_a {\cal H} F_a^c + F_a {\cal H} Q_a^c
    \\& + \bar F_a^c \left( M F_a^c + \phi_a {\cal O_{B-L}} Q_3^c + {\cal O_{B-L}} \frac{\theta_a \theta_b}{\hat M} Q_b^c + B_2 Q_a^c\right)
    \\& + \bar F_a \left( M F_a + \phi_a {\cal O_{B-L}} Q_3 + {\cal O_{B-L}} \frac{\theta_a \theta_b}{\hat M} Q_b + B_2 Q_a\right) \,,
  \end{aligned}
\end{align}
where  $ \{ Q_3, \ Q_a, \ F_a \} = (4, 2, 1,1), \;  \{ Q_3^c, \ Q^c_a, \ F_a^c \} = (\bar 4, 1, \bar 2, 1)$ with $a = 1,2$, a $D_4$ family index, ${\cal H} = (1, 2, \bar 2, 0)$ and the fields $\bar F_a,  \ \bar F_a^c$ are massive fields which are integrated out at the scale $M$ to obtain effective Yukawa matrices.  The superpotential for the neutrino sector is given by
\begin{align}
  \begin{aligned}
    \fw_{neutrino} &= \Sb^c ( \lambda_2 N_a Q^c_a + \lambda_3 N_3 Q^c_3 )
    - \frac{1}{2} \left( \lambda'_2 Y^\prime N_a N_a + \frac{\tilde \theta_a \tilde \theta_b}{\hat M} N_a N_b + \lambda'_3 Y^\prime N_3 N_3 \right)
    \\&= \sum_{i=1}^3\frac{\lambda_i^2}{2M_i}(\bar{S}^cQ_i^c)^2 \,,
  \end{aligned} \label{eq:Wnu}
\end{align}
where
\be
  M_1 = \lambda'_2 Y^\prime \, \qquad M_2 = \lambda'_2 Y^\prime + \frac{\tilde \theta_2^2}{\hat M}\, \qquad  M_3 = \lambda'_3 Y^\prime \,,
\ee
and $\widetilde{\theta}_1$ is taken to be zero.  The fields $N_3, \; N_a$ are PS singlets and $\Sb^c$ is a waterfall field, discussed later.

After expanding the waterfall field by its vacuum expectation value (vev), the last line of eq.~(\ref{eq:Wnu}) yields (with $\Sb^c \rightarrow (\sigma + i \tau + \sqrt{2} v_{PS})/\sqrt{2}$)
\be
  \frac{\lambda^2_i}{2 M_i}\lrp{\frac{\sigma+i\tau+\sqrt{2}v_{PS}}{2}}^2\bar \nu_i \bar \nu_i
  = \half M_{R_i} \bar \nu_i \bar \nu_i + \frac{h_i}{2}\lrp{\sigma+i\tau}\bar \nu_i \bar \nu_i \,,
  \label{eq:sdecay}
\ee
plus terms quadratic in $\sigma$ and $\tau$ with
\begin{align}
  M_{R_i}\equiv \frac{\lambda^2_i v^2_{PS}}{2 M_i}
  \hspace{1cm}\text{and}\hspace{1cm}
  h_i \equiv \frac{\lambda^2_i v_{PS}}{\sqrt{2}M_i} \,,
  \label{eq:MRi_hi}
\end{align}
where $\lambda_1 = \lambda_2$.

Here $Y^\prime$ is identified as one of the flavon fields. The ``right-handed" neutrino fields,  $N_a, \ N_3$ are PS singlets with charge $(1,1,1,1)$. The vev of $Y^\prime$ gives a heavy mass term for $N_a, \ N_3$ which are in turn integrated out to yield effective couplings between the waterfall field and the left-handed anti-neutrinos in $Q^c_a$ and $Q^c_3$. Similar to the waterfall field, the scalar components of $Y$ also obtain a coupling to the left-handed anti-neutrinos
\be
\frac{h_i}{2}\lrp{\frac{m}{\kappa v_{PS}}}\lrp{h+iu}\bar \nu_i \bar \nu_i
\,.
\label{eq:hdecay}
\ee

The fields $F_a, \ \bar F_a, \ F_a^c, \ \bar F_a^c$ are Froggatt-Nielson fields which are integrated out to obtain the effective Yukawa matrices.  The effective operators $\cal O_{B-L}$ and $\cal O$ are defined by
\begin{align}
  \begin{aligned}
    {\hat M^2 ({\cal O_{B-L}})^{\alpha i}}_{\beta j} \equiv& - \frac{4}{3} {\delta^i}_j \bar {S^c}^{\gamma k} \lrp{{\delta^\alpha}_\gamma {\delta^\lambda}_\beta - \frac{1}{4} {\delta^\alpha}_\beta {\delta^\lambda}_\gamma}  S^c_{\lambda k}
    \\=& {(B-L)^\alpha}_\beta {\delta^i}_j \frac{v_{PS}^2}{2} \,,
  \end{aligned}
\end{align}
and
\begin{align}
  \begin{aligned}
    {\hat M^2 {\cal O}^{\alpha i}}_{\beta j} &\equiv \bar {S^c}^{\gamma k} \left[{\delta^\alpha}_\beta {\delta^i}_j {\delta^\lambda}_\gamma {\delta^l}_k + \tilde \alpha {\delta^\lambda}_\gamma \lrp{ {\delta^i}_k {\delta^l}_j - \frac{1}{2} {\delta^i}_j {\delta^l}_k}\right.
    \\&\hspace{2.8cm} \left. -\frac{4}{3} \tilde \beta {\delta^l}_k {\delta^i}_j \lrp{{\delta^\alpha}_\gamma {\delta^\lambda}_\beta - \frac{1}{4} {\delta^\alpha}_\beta {\delta^\lambda}_\gamma} \right] S^c_{\lambda l}
    \\&= \lrs{{\mathbb{I}^{\alpha i}}_{\beta j} + \tilde \alpha {(T_{3R})^i}_j {\delta^\alpha}_\beta + \tilde \beta {(B-L)^\alpha}_\beta {\delta^i}_j } \frac{v_{PS}^2}{2}
    \\&\equiv \lrs{{\mathbb{I}^{\alpha i}}_{\beta j} + \alpha {(X)^{i \alpha}}_{j \beta} + \beta {(Y)^{i \alpha}}_{j \beta} } \frac{v_{PS}^2}{2} \,,
  \end{aligned}
\end{align}
where $X = 3 (B - L) - 4 T_{3R}$ commutes with $SU(5)$ and $Y = 2 T_{3R} + (B - L)$ is the SM hypercharge.  The Froggatt-Nielson fields $F_a, \ \bar F_a, \  F_a^c, \ \bar F_a^c$ have a mass term $M$ given by $M_0 \ {{\fo}^{\alpha i}}_{\beta j}$.  The flavon fields $\phi_a, \ \theta_a, \ \tilde \theta_a$ are doublets under $D_4$ while $B_2$ is a non-trivial $D_4$ singlet such that the product $B_2 * (x_1 y_2 - x_2 y_1)$ is $D_4$ invariant with $x_a, \ y_a$ as $D_4$ doublets.  The $D_4$ invariant product between two doublets is given by $x_a y_a \equiv  x_1 y_1 + x_2 y_2$.  All flavon fields have zero charge under \zfr.  The flavon fields $\phi_{1,2}, \ \theta_2, \ \tilde \theta_2, \ B_2$ are assumed to get non-zero vevs while all other flavon fields have zero vevs.

Note, with the given particle spectrum and \zfr charges, we have the following anomaly coefficients, \be A_{SU(4)_C-SU(4)_C-\zfrmm} = A_{SU(2)_L-SU(2)_L-\zfrmm} = A_{SU(2)_R-SU(2)_R-\zfrmm} = 1 ({\rm mod}(2)) . \ee The \zfr symmetry forbids dimension 4 and 5 operators for proton decay and also a $\mu$ term.  In addition, the \zfr anomaly can, in principle, be canceled via the Green-Schwarz mechanism, as discussed in Ref.~\cite{Lee:2010gv,Lee:2011dya,Kappl:2010yu}.  Dynamical breaking of the \zfr symmetry would then preserve an exact $R$-parity and generate a $\mu$ term, with $\mu \sim m_{3/2}$ and dimension 5 proton decay operators suppressed by  $m_{3/2}^2/\text{M}_\text{pl}$.

\subsection{Yukawa matrices}
Upon integrating out the heavy Froggatt-Nielsen fields, we obtain the effective superpotential for the low energy theory,
\begin{align}
    \fw_{LE} =  Y^u_{i j} \ q_i \ H_u \ \bar u_j + Y^d_{i j} \ q_i \ H_d \ \bar d_j + Y^e_{i j} \ \ell_i \ H_d \ \bar e_j + Y^\nu_{i j} \ \ell_i \ H_u \ \bar \nu_j + \frac{1}{2} \  M_{R_i} \bar \nu_i \ \bar \nu_i \,,
\end{align}
where $i,j=1,2,3$ and
\be
  M_{R_{1,2}} = \frac{\lambda_2^2 \ v_{PS}^2}{2 \ M_{1,2}} \,,
  \quad
  M_{R_{3}} = \frac{\lambda_3^2 \ v_{PS}^2}{2 \ M_{3}} \,.
  \label{eq:masses}
\ee
The Yukawa matrices for up-quarks, down-quarks, charged leptons and neutrinos are given by (defined in Weyl notation with doublets on the left)\footnote{\label{fn:yukawa} These Yukawa matrices are identical to those obtained previously (see Ref.~\cite{Dermisek:2005ij}) and analyzed most recently in Ref.~\cite{Anandakrishnan:2012tj,Anandakrishnan:2014nea}.}
\begin{align}
  \begin{aligned}
      Y^u =& \left(\begin{array}{ccc} 0                & \epsilon'\ \rho      & -\epsilon\ \xi \\
                                      -\epsilon'\ \rho & \tilde\epsilon\ \rho & -\epsilon      \\
                                      \epsilon\ \xi    & \epsilon             & 1
                   \end{array}\right)\;\lambda
    \\Y^d =& \left(\begin{array}{ccc} 0             & \epsilon'       & -\epsilon\ \xi\ \sigma \\
                                      -\epsilon'    & \tilde\epsilon  & -\epsilon\ \sigma      \\
                                      \epsilon\ \xi & \epsilon        & 1
                    \end{array}\right)\;\lambda
    \\Y^e =&  \left(\begin{array}{ccc} 0                         & -\epsilon'           & 3\ \epsilon \ \xi \\
                                       \epsilon'                 & 3\ \tilde\epsilon    & 3\ \epsilon       \\
                                       -3\ \epsilon\ \xi\ \sigma & -3\ \epsilon\ \sigma & 1
                    \end{array}\right)\;\lambda \,,
  \end{aligned}\label{eq:yukawaD31}
\end{align}
with
\begin{align}
  \begin{aligned}
      \xi      &=       \phi_1/\phi_2\,,             &&&\tilde\epsilon  &\propto (\theta_2/\hat M)^2\,,
    \\\epsilon &\propto -\phi_2/\hat M\,,            &&&\epsilon^\prime &\sim    ({ B_2}/M_0),
    \\\sigma   &=       \frac{1+\alpha}{1-3\alpha}\,,&&&\rho            &\sim    \beta\ll\alpha \,,
  \end{aligned}\label{eq:omegaD3}
\end{align}
and
\bea
  Y^{\nu} = \left(\begin{array}{ccc} 0                         & -\epsilon'\ \omega        & \frac{3}{2} \ \epsilon\ \xi\ \omega \\
                                     \epsilon'\ \omega         & 3\ \tilde\epsilon\ \omega & \frac{3}{2} \ \epsilon\ \omega      \\
                                     -3\ \epsilon\ \xi\ \sigma & -3\ \epsilon\ \sigma      & 1
                  \end{array} \right) \;
            \lambda
  \label{eq:yukawaD32} \,,
\eea
with $\omega=2\,\sigma/(2\,\sigma-1)$ and a Dirac neutrino mass matrix given by
\begin{equation}
  m_\nu \equiv Y^\nu\frac{v}{\sqrt{2}}\sin\beta \,.
  \label{eq:mnuD3}
\end{equation}

From eq.~(\ref{eq:yukawaD31}) and (\ref{eq:yukawaD32}), one can see that the flavor hierarchies in the Yukawa couplings are encoded in terms of the four complex parameters $\rho, \sigma, \tilde \epsilon, \xi$ and three real parameters $\epsilon, \epsilon', \lambda$.  These matrices contain 7 real parameters and 4 arbitrary phases.  While the superpotential $\fw_{PS}$ has many arbitrary parameters, the resulting effective Yukawa matrices have much fewer parameters, therefore obtaining a very predictive theory.  Also, the quark mass matrices accommodate the Georgi-Jarlskog mechanism, such that $m_\mu/m_e \approx 9 \ m_s/m_d$.  This is a result of the operator $\cal O_{B-L}$ which is assumed to have a vev in the $B - L$ direction.

\subsection{Yukawa Unification – 3rd family only}

Some of the major properties of the PS model are purely the result of analyzing only the third family.  In  particular, we consider the third family Yukawa couplings given by \be  \lambda Q_3 {\cal H} Q^c_3 . \ee  We fit the top, bottom and tau masses and three flavor violating observables \be BR(B_s \rightarrow \mu^+ \mu^-), \;\;\; BR(B_s \rightarrow K^* \ \mu^+ \ \mu^-) \;\;\; {\rm and} \;\;\; BR(B \rightarrow X_s + \gamma) . \ee  In order to calculate these processes we use the observed values of the relevant CKM mixing parameters.   For soft SUSY breaking parameters we assume universal squark and slepton masses, $m_{16}$; universal gaugino mass, $M_{1/2}$; non-universal Higgs masses, $\sqrt{m_{10}^2 \pm \Delta m^2}$; a universal A parameter, $A_0$, and $\mu$ and $\tan\beta$.  In order to fit the data we find that we are forced to the following range of parameters.
\be  A_0 \sim - 2 m_{16};  \;\;  m_{10} \sim \sqrt{2} \ m_{16}; \;\;   \mu, M_{1/2} \ll m_{16}; \;\;  m_{16}  >  10 \;\; {\rm TeV},  \ee  and \be \tan\beta \sim 50 . \ee  It is the B physics processes which force us to have heavy squarks \cite{Albrecht:2007ii}.  Note, with this range of parameters we get a bonus of an inverse scalar mass hierarchy \cite{Bagger:1999sy}.   Squarks and sleptons of the first two families have mass of order $m_{16}$, while the third generation scalars are significantly lighter.

In this range of parameters we also find the CP odd higgs boson has mass, $m_A > 1$ TeV.   Thus we are in the decoupling limit of the MSSM and the light higgs boson is necessarily very much Standard Model-like.

\subsection{Global  $\chi^2$   fits  \&  predictions}
\begin{figure}[h!]
  \centering
    \includegraphics[width=0.5\textwidth]{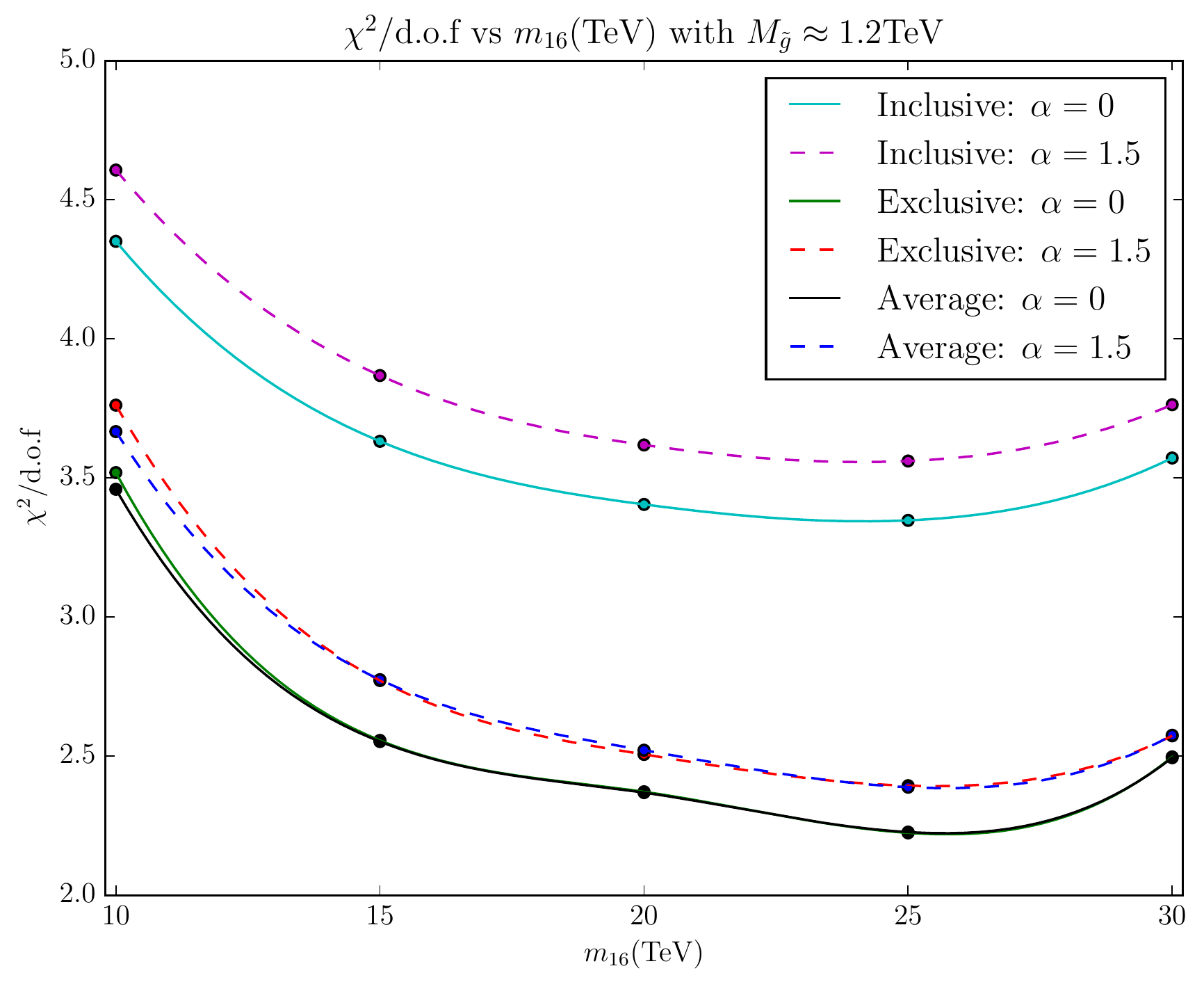}
    \caption{This plot shows the value of $\chi^2$/d.o.f as a function of $m_{16}$ for cases where the value of $|V_{ub}|$ and $|V_{cb}|$ are taken to be the inclusive values, the exclusive values, or the average of inclusive and exclusive values.  Solid lines refers to the universal boundary condition, $\alpha=0$, while dashed lines refer to the mirage boundary condition with $\alpha=1.5$.  This plot shows that our model favors the exclusive values of $|V_{ub}|$ and $|V_{cb}|$.}
    \label{fig:chi2_line_plot}
\end{figure}

As one example of the $\chi^2$ fits see Fig. \ref{fig:chi2_line_plot}.
Three of the largest pulls in $\chi^2$ were due to the observables $m_d/m_s,  \; \sin2\beta$ and $\sin^2\theta_{13}$.  Note, given our $\chi^2$ analysis we find an upper bound on the gluino mass of $M_{gluino} \leq 2.4$ TeV for values of $m_{16} \leq 30$ TeV.   Moreover, $\chi^2$ increases for larger values of $m_{16}$.

\section{Subcritical hybrid F-term inflation}
Let us now briefly review the results of Ref.~\cite{Bryant:2016tzg}.  The superpotential and K\"{a}hler potential for the inflaton sector of the model with a Pati-Salam $SU(4)_C \times SU(2)_L \times SU(2)_R$ gauge symmetry times \zfr discrete $R$ symmetry are given by
\begin{align}
  \fw_{I} &= \Phi \left( \kappa \Sb^c S^c + m_\phi Y + \frac{1}{\sqrt{2}}\alpha {\cal H} {\cal H}\right) + \lambda X\lrp{\Sb^c S^c-\frac{v^2_{PS}}{2}} + S^c \Sigma S^c + \Sb^c \Sigma \Sb^c \label{eq:superpotential}\\
  \fk &= \half(\Phi+\Phi^\dagger)^2 + (S^c)^\dagger S^c + (\Sb^c)^\dagger \Sb^c + Y^\dagger Y + X^\dagger X\lrs{1-c_X \frac{X^\dagger X}{\mpl^2}+a_X\lrp{\frac{X^\dagger X}{\mpl^2}}^2 } \,,
\end{align}
with the quantum numbers of the inflaton and waterfall superfields, respectively: $ \{\Phi = (1, 1, 1, 2), \; S^c = (\bar{4}, 1, \bar{2}, 0), \; \Sb^c = (4, 1, 2, 0) \} $.\footnote{The fields $Y$ and $Y^\prime$ can be distinguished by an additional $\mathbb{Z}_4$ symmetry where $Y$ is invariant, but $Y^\prime, \ N_a, \ N_3, \ \tilde \theta_a, \ S^c, \ \Sb^c, \ \Sigma$ have $\mathbb{Z}_4$ charges $2, \ 1, \ 1, \ 1, \ 1,  \ 3, \ 2$, respectively.}  As a consequence, the Pati-Salam gauge symmetry is broken to the Standard Model (SM) at the waterfall transition and remains this way both during inflation and afterwards.  The superfield, $\Sigma = (6,1,1,2)$, is needed to guarantee that the effective low energy theory below the PS breaking scale is just the minimal supersymmetric standard model (MSSM).

\begin{figure}[t!]
\centering
\includegraphics[width=0.9\textwidth, trim=0ex 20ex 0ex 20ex, clip]{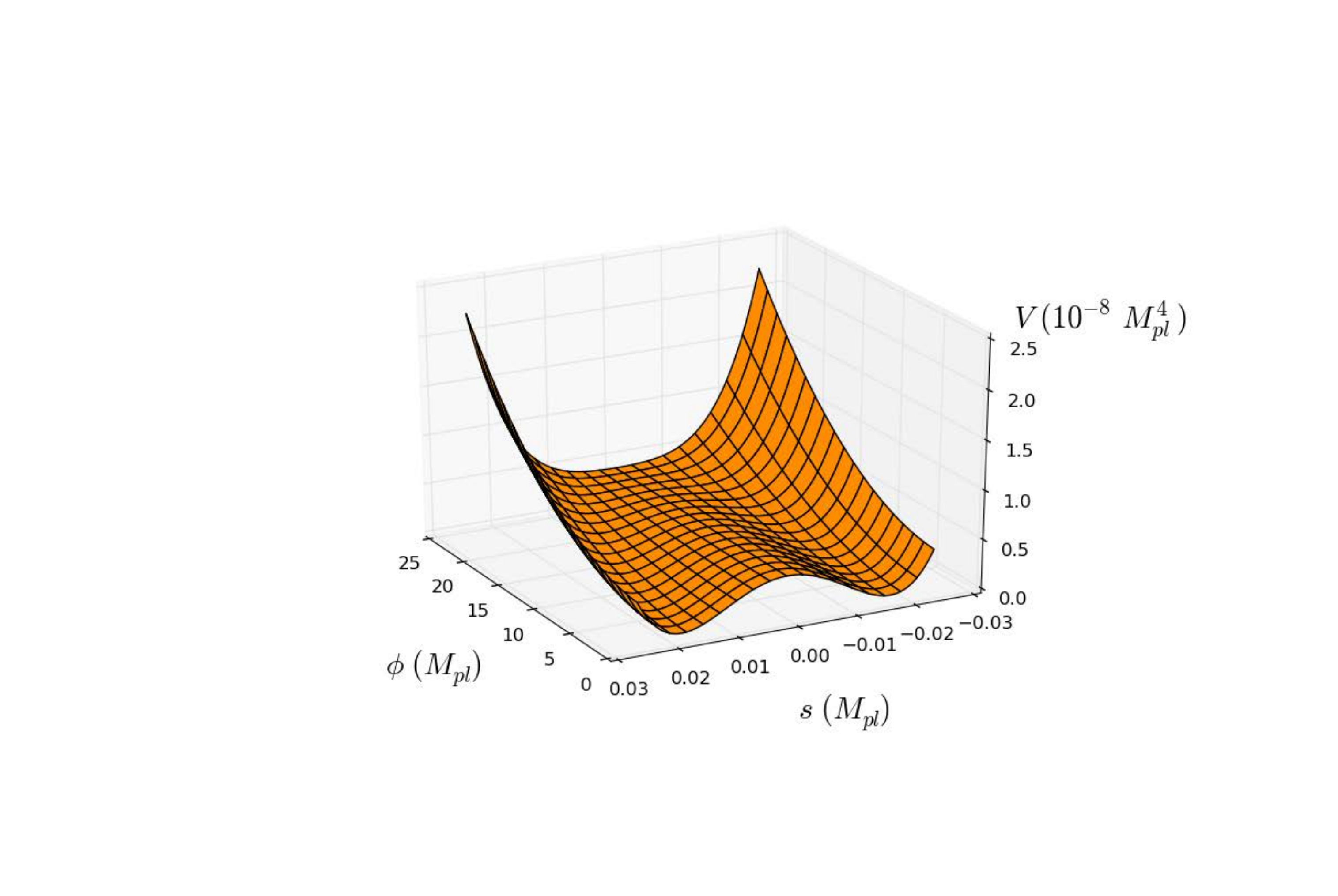}
\caption{\small The potential during inflation.  The inflaton, $\phi$, and the waterfall field, $s$, are in Planck units.}
\label{fig:3D}
\end{figure}

\begin{figure}[h!]
\centering
\includegraphics[width=0.6\textwidth, trim=0ex 0ex 0ex 0ex, clip]{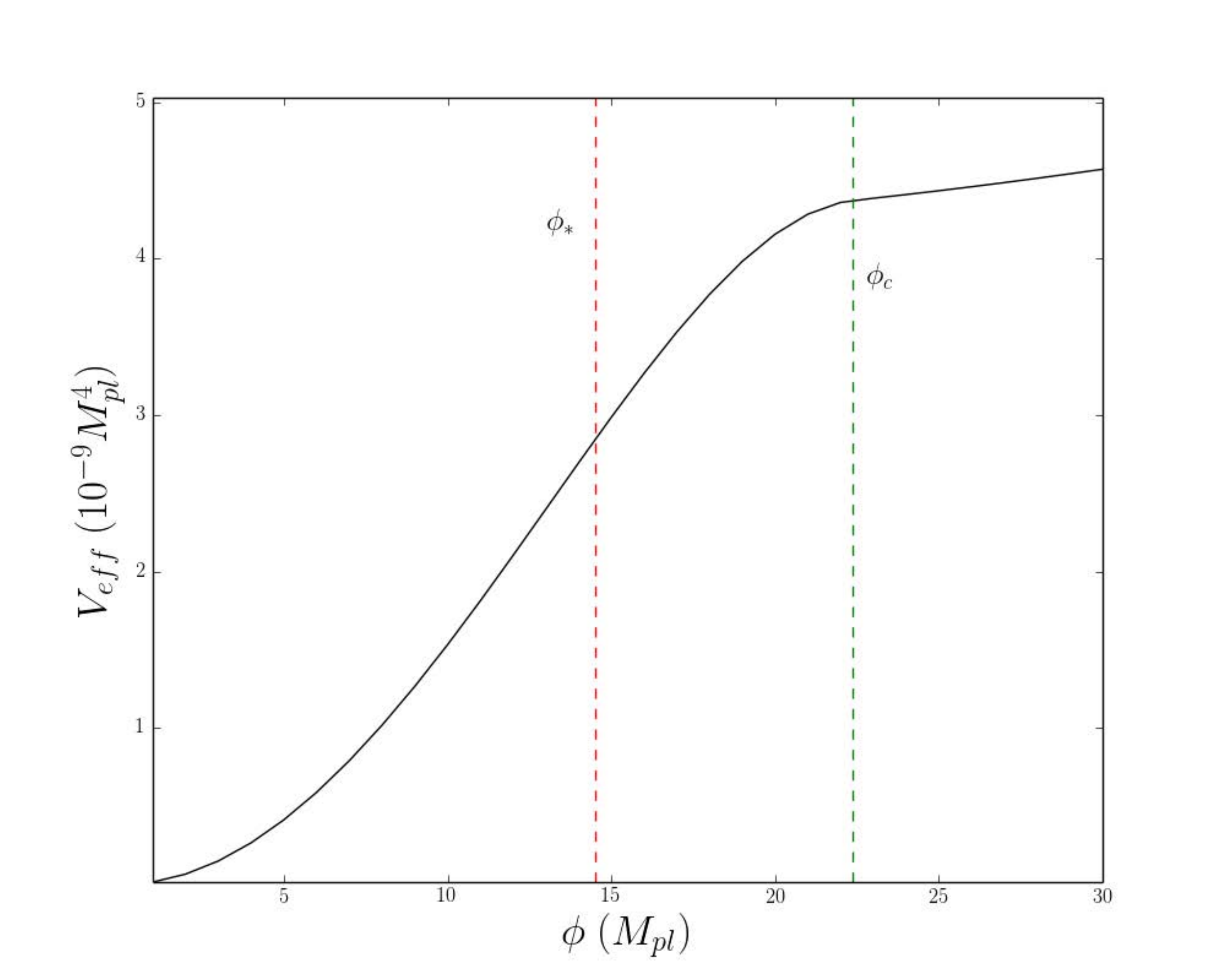}
\caption{\small The effective single-field potential during inflation. The critical point $\phi_c$  is denoted by the vertical, dashed green line. The value of $\phi$ at the start of the last 60 $e$-folds, $\phi_*$, is denoted by the vertical, dashed red line. For values of $\phi$ above $\phi_c$, the potential is given by $V_0 = \frac{\lambda^2 v^4_{PS}}{4}+\half m^2 \phi^2$.}
\label{fig:veff}
\end{figure}

\begin{figure}[h!]
\centering
\includegraphics[width=0.5\textwidth, trim=0ex 10ex 10ex 15ex, clip]{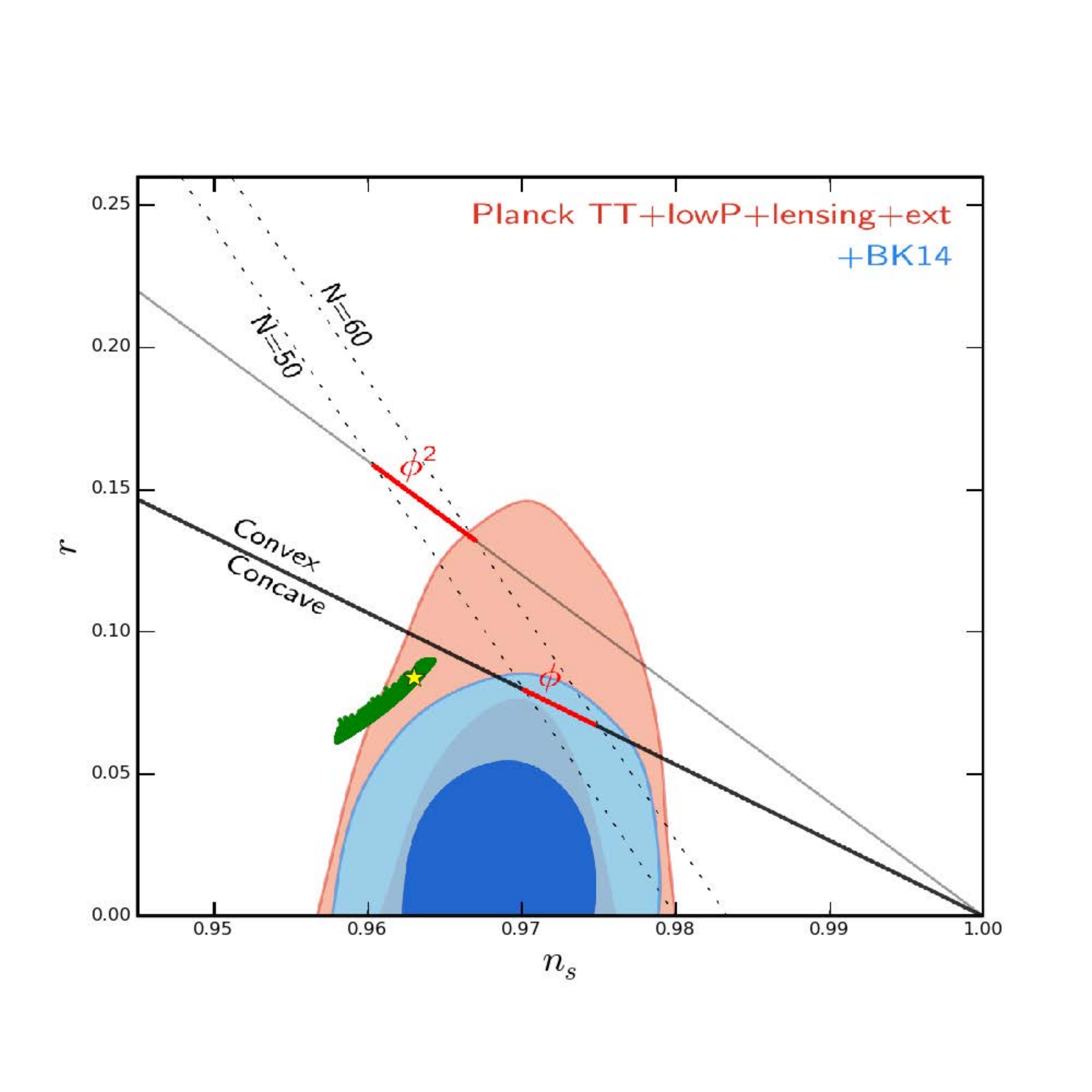}
\caption{\small The green points represent the result of our parameter scan and are overlayed on the best-fit plane found in~\cite{Array:2015xqh}. The yellow star represents our best fit point.}
\label{fig:overlay}
\end{figure}

The inflaton/waterfall potential during inflation is given in Fig. \ref{fig:3D}.  The waterfall field is initially at zero and then after the inflaton field passes the critical value, the waterfall field obtains a negative mass squared.  In this model, it is important to note that the critical value for the inflaton field is super-Planckian.   Thus the last 60 e-folds of inflation occurs at a subcritical value of the inflaton field.  Hence the name {\it subcritical hybrid inflation}.
At some point, the waterfall field rolls slowly down the side of the potential.  If we input the minimum of the potential for the waterfall field as a function of the inflaton field we obtain an effective potential for the inflaton.
\be
V_{eff}(\phi) = \frac{\lambda^2 v_{PS}^4}{2}\frac{\phi^2}{\phi_c^2}\left[\left(1+\frac{m^2}{\kappa^2 v^2_{PS}}\right)-\frac{\phi^2}{2\phi_c^2}\right]
\simeq
\frac{\lambda^2 v_{PS}^4}{2}\frac{\phi^2}{\phi_c^2}\left(1-\frac{\phi^2}{2\phi_c^2}\right)
\;,
\label{eq:veff}
\ee
where $m^2/\kappa^2 v^2_{PS}\ll 1$. The effective potential is plotted in Fig.~\ref{fig:veff}.
In Ref.~\cite{Bryant:2016tzg} we found the best fit to Bicep2-Keck-Planck data as seen in Fig. \ref{fig:overlay}.
Now we need to discuss reheating of the universe after inflation and baryogeneses via the process of leptogenesis.

\section{Reheating and Leptogenesis}

After inflation the inflaton and waterfall fields oscillate around their respective minima.  The inflaton couples directly to the Higgs via the coupling  $\alpha \ \phi \ H_u \ H_d$.   In addition to allowing the inflaton to decay directly to Higgs, the effective Higgs mass depends on the value of $\phi$.
Note, the Yukawa matrices determining the Higgs couplings to leptons are fixed by fitting low energy data.
These matrices enter the calculation of a lepton number asymmetry generated by the decay of heavy RH neutrinos and also by the decay of $H_u$ into RH neutrinos.  $H_u$ decays into RH neutrinos when its mass is greater than the RH neutrino mass. Finally, the  waterfall field also decays directly into RH neutrinos.

Higgses and RH neutrinos decay into light quarks and leptons which quickly thermalize. Thus reheating of the universe and leptogenesis occur simultaneously.  In Ref. \cite{Bryant:2016sjj} we evaluate reheating and
the baryon number of the universe relevant for big bang nucleosynthesis.

The CP asymmetry due to Higgs decay is evaluated as follows.
\begin{align}
  h_u \to
  \begin{cases}
      \dfrac{1+\epsilon_{h_i}}{2}\,\bar{\nu}_i      \ell         \to&
      \begin{cases}
          \dfrac{(1+\epsilon_{h_i})(1+\epsilon_{\bar{\nu}_i})}{4}\,h_u        \ell        \ell
        \\\dfrac{(1+\epsilon_{h_i})(1-\epsilon_{\bar{\nu}_i})}{4}\,h_u^\dagger\ell^\dagger\ell
      \end{cases}
    \\\dfrac{1-\epsilon_{h_i}}{2}\bar{\nu}_i^\dagger\ell^\dagger \to&
      \begin{cases}
          \dfrac{(1-\epsilon_{h_i})(1+\epsilon_{\bar{\nu}_i})}{4}h_u        \ell        \ell^\dagger
        \\\dfrac{(1-\epsilon_{h_i})(1-\epsilon_{\bar{\nu}_i})}{4}h_u^\dagger\ell^\dagger\ell^\dagger
      \end{cases}
  \end{cases} \,,
\end{align}
where the $\epsilon$ factors are the CP asymmetry parameters.  We see that only half of the decay channels have a net lepton asymmetry.  Hence, the final lepton asymmetry is
\begin{align}
  n_{L}
  \equiv n_{\ell}-n_{\bar{\ell}}
  = 2\frac{(1+\epsilon_{h_i})(1+\epsilon_{\bar{\nu}_i})}{4}n_{h_u} -
    2\frac{(1-\epsilon_{h_i})(1-\epsilon_{\bar{\nu}_i})}{4}n_{h_u}
  = \epsilon_{h_i}        n_{h_u}         +
    \epsilon_{\bar{\nu}_i}n_{\bar{\nu}_i}   \,.  \label{eqn:nL}
\end{align}
Finally,  the baryon number asymmetry relevant for big bang nucleosynthesis is given by $n_B = - \frac{8}{23} n_L$.

The CP asymmetry due to Higgses(Higgsinos) decay is given by
\begin{align}
  \epsilon_{h_i}
  \equiv \frac{\Gamma_{h_u^\dagger\to\bar{\nu}_i        \ell        } -
               \Gamma_{h_u        \to\bar{\nu}_i^\dagger\ell^\dagger}   }
              {\Gamma_{h_u^\dagger\to\bar{\nu}_i        \ell        } +
               \Gamma_{h_u        \to\bar{\nu}_i^\dagger\ell^\dagger}   } \,,
  \label{eq:epsi}
\end{align}
and due to right-handed (s)neutrinos decay by
\begin{align}
  \epsilon_{\bar{\nu}_i}
  \equiv \frac{\Gamma_{\bar{\nu}_i^\dagger\to\ell         h_u        } -
               \Gamma_{\bar{\nu}_i        \to\ell^\dagger h_u^\dagger}   }
              {\Gamma_{\bar{\nu}_i^\dagger\to\ell         h_u        } +
               \Gamma_{\bar{\nu}_i        \to\ell^\dagger h_u^\dagger}   } \,,
\end{align}
where the family indices of the leptons are summed.

The CP asymmetry parameters for the heavy right-handed (s)neutrinos and Higgs are given by
\begin{align}
  \begin{aligned}
    \epsilon_{\bar{\nu}_3} = \epsilon_{h_3}
    =&\, \frac{1}{8\pi}\sum_{j=1,2}\frac{\text{Im}\{[(\lambda_\nu \lambda_\nu^\dagger)_{j3}]^2\}}{(\lambda_\nu \lambda_\nu^\dagger)_{33}}f\left(\frac{m_{\bar \nu_j}}{m_{\bar \nu_3}}\right)
    \\\epsilon_{\bar{\nu}_2} = \epsilon_{h_2}
    =&\, \frac{1}{8\pi}\frac{\text{Im}\{[(\lambda_\nu \lambda_\nu^\dagger)_{12}]^2\}}{(\lambda_\nu \lambda_\nu^\dagger)_{22}}f\left(\frac{m_{\bar \nu_1}}{m_{\bar \nu_2}}\right)
    + \frac{3}{8\pi}\frac{\text{Im}[(\lambda_\nu^*M_n^{(3)}\lambda_\nu^\dagger)_{22}]}{(\lambda_\nu \lambda_\nu^\dagger)_{22}}m_{\bar \nu_2}
    \\\epsilon_{\bar{\nu}_1} = \epsilon_{h_1}
    =&\, \frac{3}{8\pi}\frac{\text{Im}[(\lambda_\nu^*M_n^{(2,3)}\lambda_\nu^\dagger)_{11}]}{(\lambda_\nu \lambda_\nu^\dagger)_{11}}m_{\bar \nu_1} \,,
  \end{aligned}
\end{align}
where we have made the assumption that the decay products are massless. The Weinberg operator $M_n^{(i)}= Y_\nu^T \frac{1}{m_{\bar \nu_i}} Y_\nu$ is
calculated by integrating out the $i^{th}$ right-handed neutrino $\bar \nu_i$ \cite{Giudice:2003jh,Buchmuller:2005eh}.
In Ref. \cite{Bryant:2016sjj} we have calculated the asymmetry parameters in our model.
We find $\epsilon_{h_3} \approx \epsilon_{\bar{\nu}_3} < 0$ and $\epsilon_{h_{1,2}} \approx \epsilon_{\bar{\nu}_{1,2}} > 0$.  Thus the decay of the heaviest right-handed neutrinos produces the correct sign of the baryon number asymmetry, while the lighter two generations produce the wrong sign.

\subsection{Instant Preheating}

The question is now, how can we obtain the correct sign of $n_B$ in our model.   It turns out that this can be accomplished due to the process
of instant preheating \cite{Felder:1998vq}. For a Lagrangian with the following term
\begin{align}
  \mathcal{L}
  \supset \frac{1}{2}\alpha^2\phi^2\chi^2 \,,
  \label{eq:broad_parametric}
\end{align}
where $\chi$ is a real scalar field, Kofman et.~al.~\cite{Kofman:1997yn} showed that when $\phi$ oscillates around $\phi=0$, $\phi$ creates $\chi$ states very efficiently at every zero-crossing.  The number density of $\chi$ created for a specific momentum $k$ is given by
\begin{align}
  n_k = \text{exp}\left(\frac{-\pi k^2}{\alpha|\dot{\phi}_0|}\right) \,,
  \label{eq:nk_created}
\end{align}
where $\dot{\phi}_0$ is the speed of $\phi$ at zero-crossing.  Hence, the number density of $\chi$ created at zero-crossing is
\begin{align}
  n_{\chi,0}
  = \int\frac{\mathrm{d}^3k}{(2\pi)^3}n_k
  = \frac{(\alpha|\dot{\phi}_0|)^{3/2}}{8\pi^3} \,,
  \label{eq:n_chi_created}
\end{align}
with a typical momentum of
\begin{align}
  k_\chi
  = \frac{1}{n_{\chi,0}}\int\frac{\mathrm{d}^3k}{(2\pi)^3}kn_k
  = \frac{2(\alpha|\dot{\phi}_0|)^{1/2}}{\pi} \,.
  \label{eq:ph}
\end{align}
The non-perturbative production of $\chi$ occurs for values of the parameter $q = \frac{\alpha^2 \phi^2_{amp}}{4 \mu^2} \gg 1$ and continues until
$q \sim 1/3$.

Instant preheating also works for fermions \cite{Greene:2000ew} (for example with ${\cal L} = \alpha \phi \ \tilde h_u \ \tilde h_d$).  At each zero crossing the inflaton loses energy and the $h, \; \tilde h$ mass increases as $\phi$ increases.  The heavy $h_u, \; \tilde h_u$ decay into heavy RH neutrinos.  The waterfall field also decays into heavy RH neutrinos.  Finally the heaviest RH neutrino is produced predominantly (as long as the Higgs mass is greater than $m_{\bar \nu_3}$), because it has the largest Yukawa coupling to the Higgs.

\begin{figure}
  \centering
    \includegraphics[height=0.35\textheight]{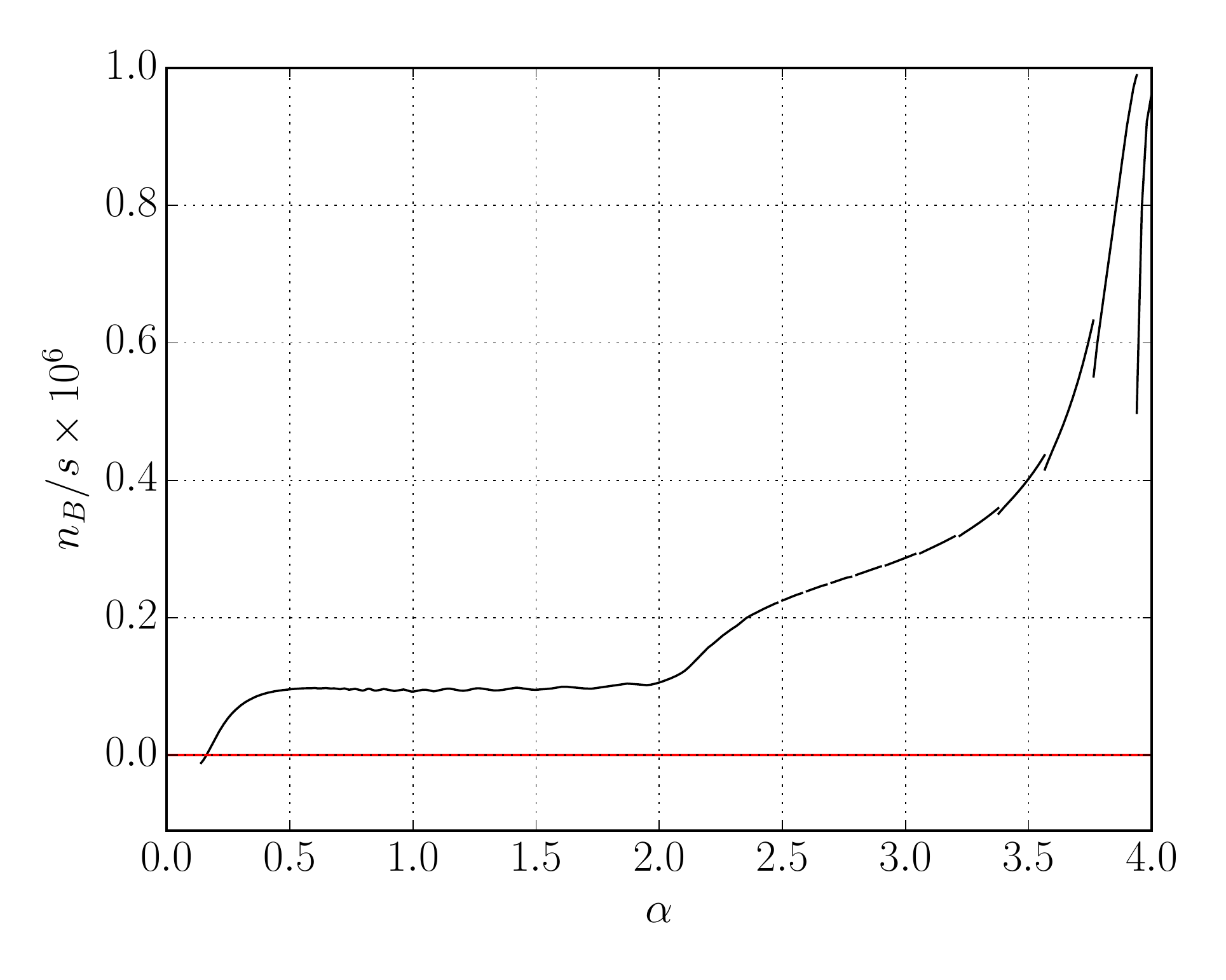}
    \includegraphics[height=0.35\textheight]{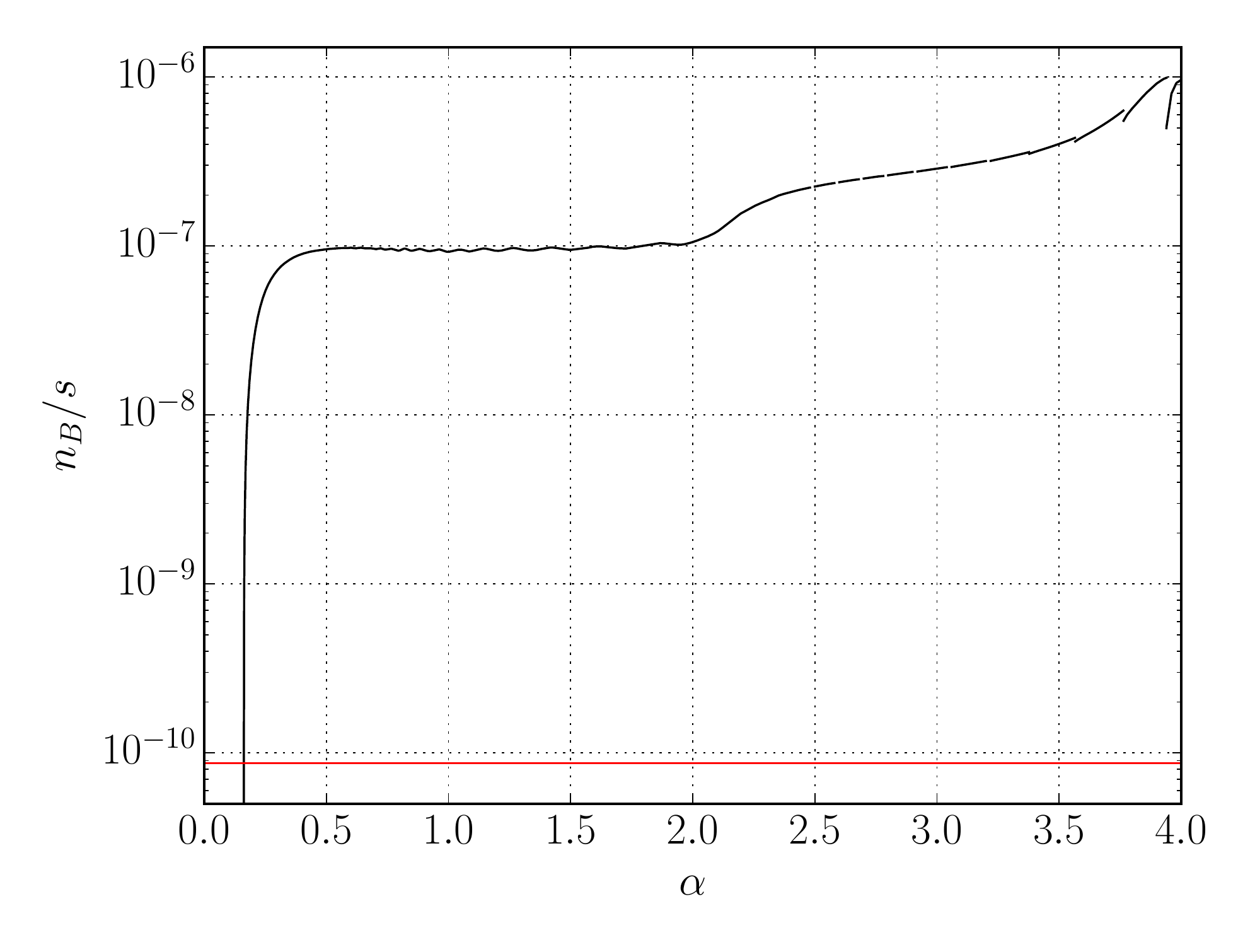}
    \caption{
   The two figures show the baryon-to-entropy ratio as a function of $\alpha$.  The left figure is in a linear-scale and the right figure has a log-scale in the $y$-axis.}
    \label{fig:nbvalpha1}
\end{figure}

\begin{figure}
  \centering
     \includegraphics[height=0.45\textheight]{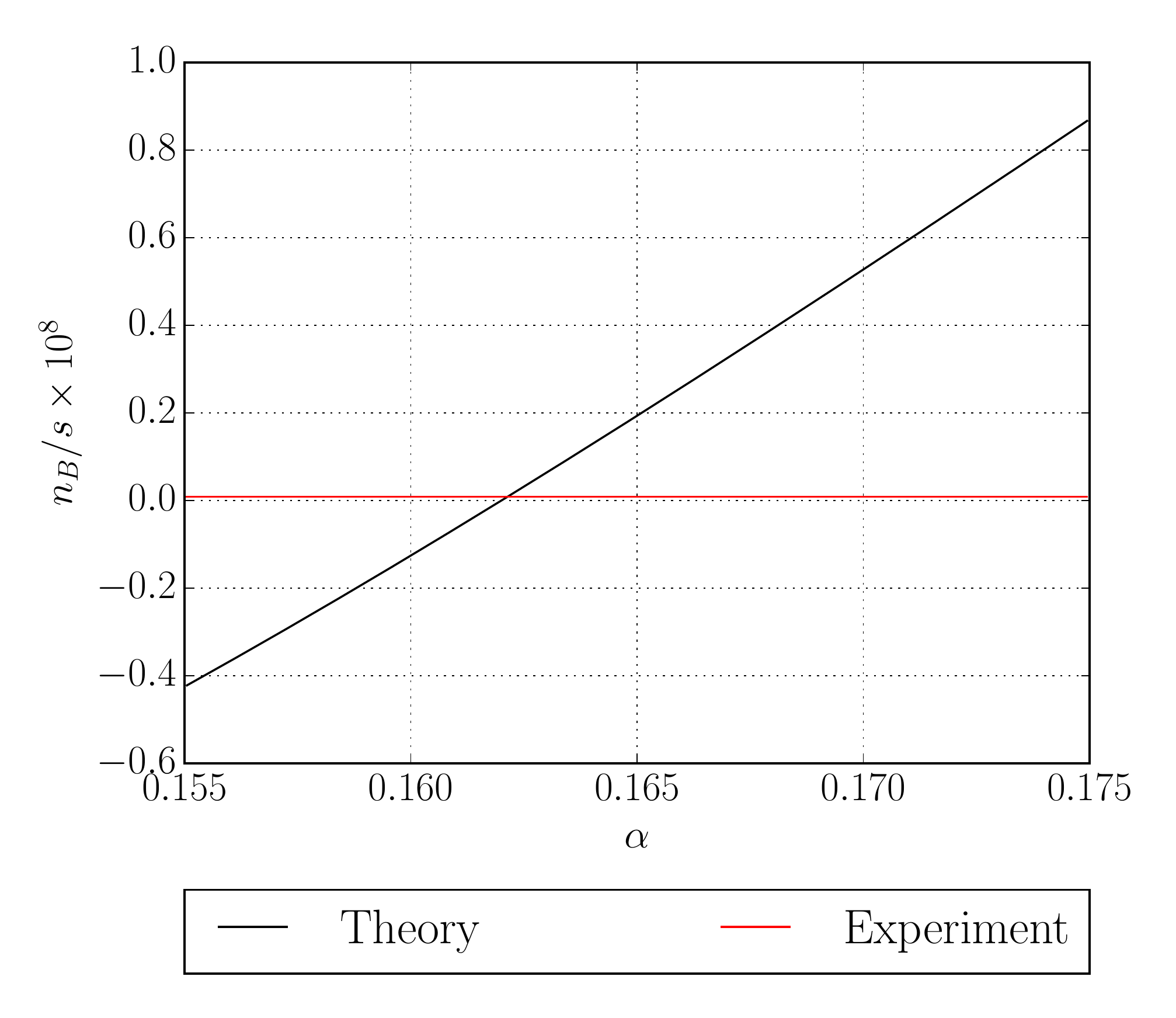}
    \caption{
    This figure shows a zoomed-in version of the baryon-to-entropy ratio as a function of $\alpha$.  In this figure the $x$-axis is zoomed-in with a linear scale.  The baryon-to-entropy ratio matches the observed value for $\alpha \sim 0.162$. }
    \label{fig:nbvalpha2}
\end{figure}

In order to evaluate the lepton number asymmetry, we have developed a set of coupled Boltzmann type evolution equations for radiation and all the Higgs and RH neutrino(s) fields.  Our evolution equations are based on the analysis used in the paper by Ahn and Kolb \cite{Ahn:2005bg}. We find the results given in Fig \ref{fig:nbvalpha1}.  A zoomed-in version of this figure is given in Fig. \ref{fig:nbvalpha2}.  The observed value of the baryon-to-entropy ratio is obtained with a value of $\alpha \sim 0.162$.
Note, that for large values of $\alpha$, the Higgs mass can be greater than the heaviest RH neutrino.  Moreover, since the decay rate is proportional to the Yukawa coupling squared, it decays predominantly to the heaviest RH neutrino.  As a result we obtain $n_B/s$ with the correct sign.   However as $\alpha$ decreases, the inflaton field spends less time in the regime where the Higgs mass is greater than the heaviest RH neutrino.  As a result $n_B/s$ eventually changes sign.   Thus we can successfully obtain the correct value for the baryon-to-entropy ratio, making use of the process of {\it instant preheating}!!

\section{Conclusions}

In this talk I have described a complete model of fermion masses and mixing angles which fits low energy data.  The largest discrepancy comes from fitting the up and down quark masses.  Since it is a SUSY GUT we are also constrained by the lack of direct evidence for supersymmetry.   We note that the best fits require the gluino mass to be less than about 2.4 TeV.   We have then extended the model to include an inflationary sector.  We are able to fit the latest Bicep2-Keck-Planck data.   Of course, following inflation the universe must reheat and we must be able to obtain a net baryon-to-entropy ratio consistent with the data.   Since all the Yukawa couplings, including the CP violating phases, and the RH neutrino masses are determined by fitting the low energy data, to be able to now fit the observed baryon-to-entropy ratio (with the correct sign) is not a given.   However we have shown that by using the process of instant preheating and a free parameter, $\alpha$, we are able to successfully fit the observed baryon-to-entropy ratio.   In the future we need to consider several other cosmological issues, such as the possible gravitino and moduli problems and dark matter.  Already we know that since our reheat temperature is of order $10^{14}$ GeV,  our SUSY LSP must necessarily be very light even though the gravitino mass is of order 30 TeV in our model.

\section{ACKNOWLEDGMENTS}
I received partial support from a Department of Energy grant DE-SC0011726.  I am also greatful for support from CETUP* 2016 Conference where this work was presented.


\vspace{.5in}

\nocite{*}
\begin{thebibliography}{999}


\bibitem{Bryant:2016tzg}
  B.~C.~Bryant and S.~Raby,
  Phys.\ Rev.\ D {\bf 93}, no. 9, 095003 (2016)
  doi:10.1103/PhysRevD.93.095003
  [arXiv:1601.03749 [hep-ph]].

\bibitem{Anandakrishnan:2014nea}
  A.~Anandakrishnan, B.~C.~Bryant and S.~Raby,
  Phys.\ Rev.\ D {\bf 90}, no. 1, 015030 (2014)
  doi:10.1103/PhysRevD.90.015030
  [arXiv:1404.5628 [hep-ph]].

\bibitem{Anandakrishnan:2012tj}
  A.~Anandakrishnan, S.~Raby and A.~Wingerter,
  Phys.\ Rev.\ D {\bf 87}, no. 5, 055005 (2013)
  doi:10.1103/PhysRevD.87.055005
  [arXiv:1212.0542 [hep-ph]].

\bibitem{Poh:2015wta}
  Z.~Poh and S.~Raby,
  Phys.\ Rev.\ D {\bf 92}, no. 1, 015017 (2015)
  doi:10.1103/PhysRevD.92.015017
  [arXiv:1505.00264 [hep-ph]].

\bibitem{Dermisek:2005ij}
  R.~Dermisek and S.~Raby,
  Phys.\ Lett.\ B {\bf 622}, 327 (2005)
  doi:10.1016/j.physletb.2005.07.018
  [hep-ph/0507045].

\bibitem{Lee:2010gv}
  H.~M.~Lee, S.~Raby, M.~Ratz, G.~G.~Ross, R.~Schieren, K.~Schmidt-Hoberg and P.~K.~S.~Vaudrevange,
  Phys.\ Lett.\ B {\bf 694}, 491 (2011)
  doi:10.1016/j.physletb.2010.10.038
  [arXiv:1009.0905 [hep-ph]].

\bibitem{Lee:2011dya}
  H.~M.~Lee, S.~Raby, M.~Ratz, G.~G.~Ross, R.~Schieren, K.~Schmidt-Hoberg and P.~K.~S.~Vaudrevange,
  Nucl.\ Phys.\ B {\bf 850}, 1 (2011)
  doi:10.1016/j.nuclphysb.2011.04.009
  [arXiv:1102.3595 [hep-ph]].

\bibitem{Kappl:2010yu}
  R.~Kappl, B.~Petersen, S.~Raby, M.~Ratz, R.~Schieren and P.~K.~S.~Vaudrevange,
  Nucl.\ Phys.\ B {\bf 847}, 325 (2011)
  doi:10.1016/j.nuclphysb.2011.01.032
  [arXiv:1012.4574 [hep-th]].

\bibitem{Albrecht:2007ii}
  M.~Albrecht, W.~Altmannshofer, A.~J.~Buras, D.~Guadagnoli and D.~M.~Straub,
  JHEP {\bf 0710}, 055 (2007)
  doi:10.1088/1126-6708/2007/10/055
  [arXiv:0707.3954 [hep-ph]].

\bibitem{Bagger:1999sy}
  J.~A.~Bagger, J.~L.~Feng, N.~Polonsky and R.~J.~Zhang,
  Phys.\ Lett.\ B {\bf 473}, 264 (2000)
  doi:10.1016/S0370-2693(99)01501-4
  [hep-ph/9911255].

\bibitem{Bryant:2016sjj}
  B.~C.~Bryant, Z.~Poh and S.~Raby,
  arXiv:1612.04382 [hep-ph].

\bibitem{Array:2015xqh}
  P.~A.~R.~Ade {\it et al.} [BICEP2 and Keck Array Collaborations],
  Phys.\ Rev.\ Lett.\  {\bf 116}, 031302 (2016)
  doi:10.1103/PhysRevLett.116.031302
  [arXiv:1510.09217 [astro-ph.CO]].


\bibitem{Giudice:2003jh}
  G.~F.~Giudice, A.~Notari, M.~Raidal, A.~Riotto and A.~Strumia,
  Nucl.\ Phys.\ B {\bf 685}, 89 (2004)
  doi:10.1016/j.nuclphysb.2004.02.019
  [hep-ph/0310123].

\bibitem{Buchmuller:2005eh}
  W.~Buchmuller, R.~D.~Peccei and T.~Yanagida,
  Ann.\ Rev.\ Nucl.\ Part.\ Sci.\  {\bf 55}, 311 (2005)
  doi:10.1146/annurev.nucl.55.090704.151558
  [hep-ph/0502169].

\bibitem{Felder:1998vq}
  G.~N.~Felder, L.~Kofman and A.~D.~Linde,
  Phys.\ Rev.\ D {\bf 59}, 123523 (1999)
  doi:10.1103/PhysRevD.59.123523
  [hep-ph/9812289].

\bibitem{Kofman:1997yn}
  L.~Kofman, A.~D.~Linde and A.~A.~Starobinsky,
  Phys.\ Rev.\ D {\bf 56}, 3258 (1997)
  doi:10.1103/PhysRevD.56.3258
  [hep-ph/9704452].

\bibitem{Greene:2000ew}
  P.~B.~Greene and L.~Kofman,
  Phys.\ Rev.\ D {\bf 62}, 123516 (2000)
  doi:10.1103/PhysRevD.62.123516
  [hep-ph/0003018].

\bibitem{Ahn:2005bg}
  E.~J.~Ahn and E.~W.~Kolb,
  Phys.\ Rev.\ D {\bf 74}, 103503 (2006)
  doi:10.1103/PhysRevD.74.103503
  [astro-ph/0508399].

\end{thebibliography}
\end{document}